\documentclass[aps,pra,reprint,superscriptaddress,longbibliography,floatfix]{revtex4-2}
\usepackage[T1]{fontenc}
\usepackage{mathptmx,amsmath,amssymb,bm,graphicx,booktabs,placeins}
\usepackage[colorlinks=true,linkcolor=blue,citecolor=blue,urlcolor=blue]{hyperref}
\hypersetup{pdftitle={Strong quantum Mpemba effect in the preparation of a Rydberg many-body steady state},pdfauthor={Junxiao Chen, Xiangyu Yang, Yiming Liu, Mingdi Xu, Lei Pan}}
\begin{document}
\title{Strong quantum Mpemba effect in the preparation of a Rydberg many-body steady state}
\author{Junxiao Chen}
\affiliation{School of Physics, Nankai University, Tianjin 300071, China}
\author{Xiangyu Yang}
\affiliation{School of Physics, Nankai University, Tianjin 300071, China}
\author{Yiming Liu}
\affiliation{School of Physics, Nankai University, Tianjin 300071, China}
\author{Mingdi Xu}
\email{2120230195@mail.nankai.edu.cn}
\affiliation{School of Physics, Nankai University, Tianjin 300071, China}
\author{Lei Pan}
\email{panlei@nankai.edu.cn}
\affiliation{School of Physics, Nankai University, Tianjin 300071, China}
\date{November 30, 2026}
\begin{abstract}
We find a strong quantum Mpemba effect in a driven Rydberg array. At pump strength $\kappa/g=0.1$, the all-ground array starts farther from the correlated steady state than a Gibbs reference at $\beta g=1$, overtakes it, and relaxes at a larger asymptotic rate. At resonance, excitation parity combined with complex conjugation confines the all-ground trajectory to the even sector. The Gibbs reference populates a slow odd mode. A $\pi/2$ pulse on a randomly chosen even or odd sublattice also populates this mode. Reversing the pulse phase preserves the Rydberg population and reverses the transverse spin throughout the evolution. Subtracting the two readouts isolates the slow coherence. A readout combining population and transverse spin settles sooner for the all-ground array than for either pulse phase.
\end{abstract}
\maketitle
\section{Introduction}

Rydberg arrays can be assembled atom by atom, driven coherently, and imaged with single-site resolution~\cite{Saffman2010,Browaeys2020,Labuhn2016,Bernien2017}. Combined with reservoir engineering, these capabilities give access to correlated many-body states~\cite{Diehl2008,Kraus2008,Verstraete2009,Muller2012,Harrington2022}. Experiments on dissipative quantum gases have resolved the relaxation of coherence~\cite{Bouganne2020} and the influence of many-body correlations on loss~\cite{Zhao2025}. In Rydberg arrays, exciting one atom shifts its neighbors out of resonance, giving rise to the Rydberg blockade~\cite{Petrosyan2013,LesanovskyGarrahan2013}. Under excitation pumping, this blockade gives initial excitations a dual role: they can bring the excitation density closer to its stationary value while preventing further excitation of neighboring sites. To accelerate the dissipative preparation of the correlated steady state, should one seed the array with excitations or start with every atom in its ground state?

The Mpemba effect, first studied in classical cooling and heating~\cite{Mpemba1969,Lasanta2017,Kumar2020,Kumar2022}, offers a reason why the more distant preparation can arrive first. Two initial states coupled to the same reservoir can have different overlaps with its slowest decay modes~\cite{LuRaz2017,Klich2019}. If one preparation avoids a slow mode and acquires a larger asymptotic relaxation rate, it can exhibit a strong quantum Mpemba effect~\cite{Carollo2021,Kochsiek2022}. Entanglement asymmetry reveals Mpemba crossings during local symmetry restoration in spin chains~\cite{Murciano2024,Rylands2024,Chalas2024,Joshi2024}, two-dimensional free-fermion lattices~\cite{YamashikaFermions2024}, and charge-conserving random Clifford circuits~\cite{YamashikaClifford2026}. Crossings also occur in the quantum Fisher information for spin rotations within a subsystem after a quench~\cite{YamashikaFisher2025}. In open systems, examples range from conductors and oscillators~\cite{Chatterjee2023,Nava2024,LonghiOscillator2025} to dissipative lattices~\cite{Dong2025,Westhoff2025}. For reviews see Refs.~\cite{AresReview2025,YuReview2025}. Trapped-ion experiments have observed inverse and strong Mpemba relaxation~\cite{Shapira2024,Zhang2025}. In a photonic experiment, Zhan \emph{et al.} eliminated slow coherence modes by preparing a qubit diagonal in its energy basis~\cite{ZhanXue2026}. Related work has explored mode selection in thermalizing spins~\cite{CaldasPires2026} and in nuclear-spin relaxation~\cite{ChatterjeeNMR2025}. In a many-body array the full-state distance reveals which preparation arrives first. Detecting this rate difference locally requires a measured observable with appreciable overlap with the slow mode.

In our earlier work on dephased spin chains, magnetization conservation and Hamiltonian symmetries restrict the modes that participate in heating~\cite{Wei2026,WeiComb2026,SongLongXY2026}. Coherence order plays a related role in ferromagnetic $SU(N)$ models~\cite{LouSUN2026}. For a dephased PXP chain, the Hamiltonian itself is a left decay eigenoperator, and its initial expectation fixes the amplitude of that mode~\cite{XuPXP2026}. Here both the coherent drive and the uniform pump change the Rydberg excitation number. In the steady state, a nonzero local coherence is required to balance the population increase induced by incoherent pumping. During relaxation, the energy couples to other observables, and its initial expectation value alone does not determine the amplitude of the slow mode.

At $\kappa/g=0.1$ the all-ground array starts farther from the steady state and relaxes faster than either the Gibbs reference at $\beta g=1$ or a reference made by pulsing one of the two sublattices with equal probability. The steady state has finite excitation density and anticorrelated nearest-neighbor occupations. Both references populate a slow odd mode. Excitation density couples only to even modes, and transverse spin detects the odd mode. A readout combining the two settles sooner for the all-ground array at the tolerances studied. In an XXZ chain with endpoint loss, coherence between reflection-related configurations speeds up relaxation at fixed initial magnetization. Mixing vacuum into the XXZ reference gives unequal initial trace distances.

\section{Rydberg array with excitation pumping}
\label{sec:model}

Each atom in the array has two internal states, a ground state $|g\rangle_j$ and a Rydberg state $|r\rangle_j$~\cite{Saffman2010,Browaeys2020}. In the frame rotating with the laser, the Hamiltonian in angular-frequency units and in the rotating-wave approximation reads
\begin{equation}
 H_{\rm R}=\sum_j\left(\frac{\Omega_j}{2}\sigma_j^x-\Delta_j n_j\right)
 +\sum_{i<j}V_{ij}n_i n_j.
 \label{eq:rydberg_array}
\end{equation}
Here $\sigma_j^+=|r\rangle_j\langle g|_j$,
$\sigma_j^-=|g\rangle_j\langle r|_j$,
$\sigma_j^x=\sigma_j^++\sigma_j^-$,
and $n_j=|r\rangle_j\langle r|_j$. The laser has real Rabi frequency $\Omega_j$ and detuning $\Delta_j$. Two excited atoms at positions $\boldsymbol r_i$ and $\boldsymbol r_j$ interact through the van der Waals potential $V_{ij}=C_6/|\boldsymbol r_i-\boldsymbol r_j|^6$.

In a chain with $|V_{j,j+2}|\ll\Omega\ll|V_{j,j+1}|$, exciting an atom shifts its nearest neighbors out of resonance~\cite{Urban2009,Gaetan2009}. The allowed configurations span the blockaded Hilbert space $\mathcal B_L$ and satisfy $n_jn_{j+1}=0$. With $P_j=1-n_j$, the projected spin flip takes the form $P_{j-1}\sigma_j^xP_{j+1}$~\cite{Lesanovsky2011,Pan2022,Cheng2023}. We take a uniform resonant drive, $\Omega_j=\Omega$ and $\Delta_j=0$, and neglect the residual interaction tail. The projected Hamiltonian is then
\begin{equation}
 H=g\sum_{j=1}^L P_{j-1}\sigma_j^xP_{j+1},\qquad g=\Omega/2.
 \label{eq:pxp_hamiltonian}
\end{equation}
The numerical calculations use periodic boundary conditions.

In the presence of incoherent excitation pumping, the density matrix obeys the Markovian master equation~\cite{Lindblad1976,GKS1976}
\begin{equation}
 \begin{aligned}
 \frac{d\rho}{dt}&=\mathcal L(\rho)=-i[H,\rho]\\
 &\quad+\sum_j\left[J_j\rho J_j^\dagger
 -\frac12\{J_j^\dagger J_j,\rho\}\right].
 \end{aligned}
 \label{eq:pump_master}
\end{equation}
A uniform pump of rate $\kappa$ excites a ground-state atom when both nearest neighbors are in $|g\rangle$. Its jump operators are
\begin{equation}
 J_j=\sqrt\kappa\,P_{j-1}\sigma_j^+P_{j+1}.
 \label{eq:pump_jumps}
\end{equation}
In an optical implementation the excitation linewidth and the dynamics of the auxiliary levels determine how well the pump resolves the blockade shift~\cite{Marcuzzi2014,Levi2016}.

The excitation number $N_r=\sum_j n_j$ is not conserved by either the coherent drive or the pump. The generator respects translations and reflections of the ring and has a further symmetry at resonance. Excitation parity $\Pi=(-1)^{N_r}$ changes the sign of $H$. Combined with complex conjugation in the configuration basis, it defines the real-linear map $\Theta(X)=\Pi X^*\Pi$, which commutes with $\mathcal L$ on Hermitian operators.

The target of the preparation is the stationary density matrix $\rho_{\rm ss}$ of the pumped array. We obtain it by solving $\mathcal L(\rho_{\rm ss})=0$ in the blockade basis, with $\rho_{\rm ss}\ge0$ and $\operatorname{Tr}\rho_{\rm ss}=1$. In this basis $\rho_{\rm ss}=\sum_{a,b}\rho^{\rm ss}_{ab}|a\rangle\langle b|$, where $a$ and $b$ label allowed configurations. The pump is nonunital and sustains a finite excitation density $\bar n_{\rm ss}=\langle N_r\rangle_{\rm ss}/L$. Blockade makes neighboring occupations anticorrelated, since $\langle n_j n_{j+1}\rangle_{\rm ss}=0$. Coherent de-excitation balances the pump inflow, which requires off-diagonal coherence even in the one-site state (Appendix~\ref{app:steady}).

Let $R_\nu$ and $L_\nu$ satisfy $\mathcal L R_\nu=\lambda_\nu R_\nu$ and $\mathcal L^\dagger L_\nu=\lambda_\nu^*L_\nu$, with $\operatorname{Tr}(L_\mu^\dagger R_\nu)=\delta_{\mu\nu}$. On a diagonalizable relaxing subspace,
\begin{equation}
 \begin{aligned}
 \rho(t)&=\rho_{\rm ss}+\sum_{\nu\ne0}c_\nu e^{\lambda_\nu t}R_\nu,\\
 c_\nu&=\operatorname{Tr}[L_\nu^\dagger(\rho(0)-\rho_{\rm ss})].
 \end{aligned}
 \label{eq:spectral_expansion}
\end{equation}
For preparation $i$, the slowest populated mode sets the decay rate $\Gamma_i=\min_{\nu\ne0:\,c_{\nu,i}\ne0}(-\operatorname{Re}\lambda_\nu)$~\cite{Carollo2021}. Jordan blocks can multiply the decaying exponentials by powers of time and preserve this exponential rate.

Following Ref.~\cite{Wei2026}, we use the squared Hilbert--Schmidt distance
\begin{equation}
 D_i(t)=\operatorname{Tr}\{[\rho_i(t)-\rho_{\rm ss}]^2\}.
 \label{eq:distance}
\end{equation}
The long-time exponent of $D_i$ is $2\Gamma_i$. A preparation $F$ that starts farther from the target than a reference $S$, $D_F(0)>D_S(0)$, exhibits a strong quantum Mpemba effect when $\eta=\Gamma_F/\Gamma_S>1$. The distances then cross and $\lim_{t\to\infty}D_F(t)/D_S(t)=0$. This criterion uses only the initial ordering and the late-time decay, so it applies also to nonunital dynamics for which the squared Hilbert--Schmidt distance can be nonmonotonic. On a finite chain of Hilbert-space dimension $d_L$, the trace distance satisfies $\sqrt{D_i}/2\le D_{{\rm tr},i}\le\sqrt{d_LD_i}/2$ and therefore has exponential rate $\Gamma_i$. If $D_{\mathrm{tr},F}(0)>D_{\mathrm{tr},S}(0)$ also holds, the trace-distance curves must cross.

\section{Strong Mpemba relaxation in the Rydberg array}

\begin{figure*}[t]
\includegraphics[width=.98\textwidth]{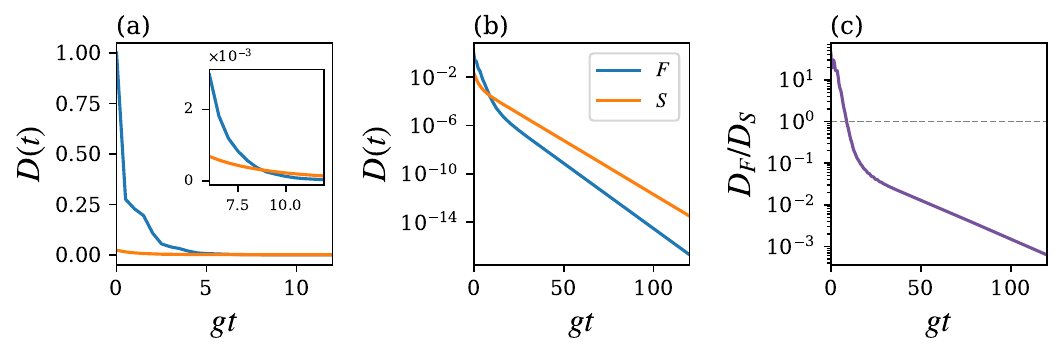}
\caption{Relaxation of a pumped ring with $L=18$, $\kappa/g=0.1$, and $\beta g=1$. Blue denotes the all-ground state $F$ and orange the Gibbs reference $S$. (a) Squared Hilbert--Schmidt distances from the same steady state. The initial distances are $1.000$ and $0.02278$, respectively. The inset resolves the crossing near $gt=8.8$. (b) The same distances over the full evolution. (c) The distance ratio $D_F/D_S$ reaches $6.3\times10^{-4}$ at $gt=120$. The dashed line marks $D_F/D_S=1$.}\label{fig:dyn}
\end{figure*}

We switch on the same excitation pump for two preparations of the periodic array. The first has every atom in its ground state, $\rho_F=|g\cdots g\rangle\langle g\cdots g|$. The second, $\rho_S=e^{-\beta H}/Z$, is a Gibbs state of the coherent PXP Hamiltonian. The all-ground array starts farther from the pumped steady state in both squared Hilbert--Schmidt distance and trace distance. It nevertheless overtakes the Gibbs reference [Fig.~\ref{fig:dyn}(a)]. At late times the all-ground distance has the steeper slope on a semilogarithmic plot. Their distance ratio $D_F/D_S$ falls by several orders of magnitude [Figs.~\ref{fig:dyn}(b,c)]. The norm bounds in Sec.~\ref{sec:model} imply that $\Gamma_F>\Gamma_S$ also holds for the trace distance.

The trajectories follow the evolution of $\rho_i-\rho_{\rm ss}$ in all symmetry blocks populated by each preparation, including recycling from every allowed jump~\cite{Dalibard1992,SanchezMunoz2019}. Appendix~\ref{app:num} describes the propagation and independent spectral checks.

\section{Symmetry selection of the slow mode}

In the configuration basis, complex conjugation reverses the sign of the Hamiltonian term in the generator. Excitation parity $\Pi$ reverses it again, while the sign acquired by each jump cancels in the dissipator. The combined operation $\Theta(X)=\Pi X^*\Pi$ therefore splits the real space of Hermitian deviations into invariant even and odd sectors. Translation and reflection further restrict the blocks accessible to the two preparations~\cite{BucaProsen2012,AlbertJiang2014,Minganti2018}.

\begin{figure*}[t]
\includegraphics[width=.98\textwidth]{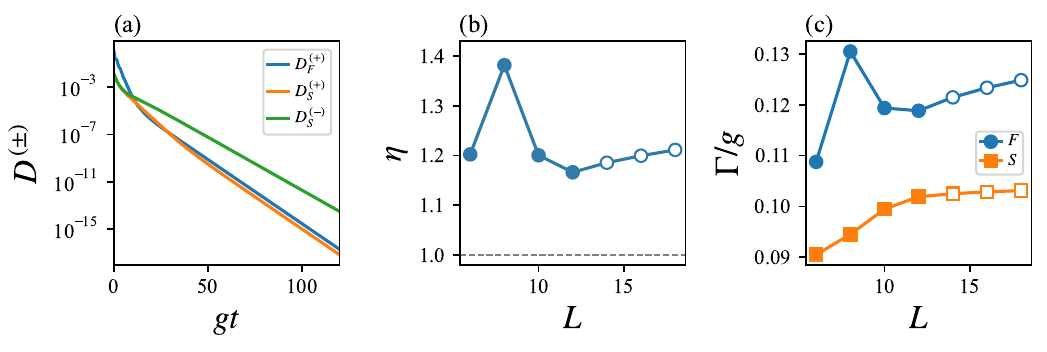}
\caption{Symmetry sectors and relaxation rates at $\kappa/g=0.1$ and $\beta g=1$. (a) Even ($+$) and odd ($-$) contributions to the squared distance at $L=18$. The all-ground state has $D_F^{(-)}=0$, while the late Gibbs tail lies in the odd sector. (b) Rate ratio $\eta=\Gamma_F/\Gamma_S$ for $L=6,8,\ldots,18$. The dashed line marks $\eta=1$. In (b,c), filled symbols denote complete accessible spectra through $L=12$ and open symbols denote sparse spectral calculations. (c) Rates of the density-operator deviations. At $L=18$, sparse diagonalization gives $\Gamma_F/g=0.1249$ and $\Gamma_S/g=0.1031$, giving $\eta_{18}\simeq1.21$. Squared distances decay with twice these exponents. Lines in (b,c) guide the eye.}\label{fig:mechanism}
\end{figure*}

The all-ground state and the stationary state are both even, so $\rho_F(t)-\rho_{\rm ss}$ stays even. For the Gibbs state, $\Theta(\rho_S)=e^{\beta H}/Z$ because the chiral symmetry gives identical partition functions at $\pm\beta$. The odd part of its initial deviation is therefore $-\sinh(\beta H)/Z$. The two parities are orthogonal in the Hilbert--Schmidt inner product. Writing $\delta\rho_i^{(\pm)}=(\delta\rho_i\pm\Theta(\delta\rho_i))/2$ and $D_i^{(\pm)}=\operatorname{Tr}[(\delta\rho_i^{(\pm)})^2]$, we obtain $D_F=D_F^{(+)}$ and $D_S=D_S^{(+)}+D_S^{(-)}$. In Fig.~\ref{fig:mechanism}(a), the odd component comes to dominate the Gibbs distance. This slow mode is inaccessible to the all-ground array. The dependence on the initial overlap also arises in studies of typical states and of other symmetries~\cite{BaoTypicality2026,Bhore2025}.

The leading populated eigenmodes in the two trajectories give $\Gamma_F>\Gamma_S$, in agreement with independent fits to their late-time slopes. The difference sets the slope of the log distance ratio in Fig.~\ref{fig:dyn}(c),
\begin{equation}
 \lim_{t\to\infty}\frac{1}{2t}\log\!\frac{D_F(t)}{D_S(t)}
 =-(\Gamma_F-\Gamma_S).
 \label{eq:rates18}
\end{equation}
Figures~\ref{fig:mechanism}(b,c) show that this rate ordering persists for all system sizes studied, up to eighteen sites. The smaller rings admit complete diagonalization in the accessible blocks. For the larger rings we obtain the leading rates and overlaps from sparse spectral calculations (Appendix~\ref{app:num}).

The odd sector has a direct physical interpretation. In the dephased PXP chain, the energy $H$ is an exact left eigenoperator~\cite{XuPXP2026}. For the pumped array, writing $h_j=gP_{j-1}\sigma_j^xP_{j+1}$ gives
\begin{equation}
 \mathcal L^\dagger H=-\frac{\kappa}{2}\sum_jh_j(1+P_{j-2}+P_{j+2}).
\end{equation}
The neighboring projectors couple the energy to other odd observables. By symmetry, their expectation values vanish throughout the evolution of an even preparation but can be nonzero for reference states with odd-sector coherence. The coherent PXP chain and the behavior of its scar states under noise are discussed in Refs.~\cite{Turner2018,Serbyn2021,Desaules2022,Ma2025}.

Removing the neighbor projectors from both drive and pump gives an independent-atom comparison. The Bloch equations already predict a faster rate for the all-ground preparation (Appendix~\ref{app:controls}). Blockade increases both populated rates at the parameters of Fig.~\ref{fig:mechanism}, reduces their ratio, and produces the spatially correlated stationary state of Sec.~\ref{sec:model}.

At $\kappa/g=0.1$, both populated rates in Fig.~\ref{fig:mechanism}(c) are of order $\kappa$ and rise from $L=12$ to $18$. The rate ratio has a local maximum at $L=8$.

For the $L=6$ and $8$ rings at $\beta g=1$, increasing the pump strength brings the leading even decay rate below the leading odd rate. At the stronger pump strengths sampled, the all-ground, Gibbs, and pulsed preparations share this even-mode rate. At fixed $\kappa/g=0.1$, a sufficiently cold Gibbs reference starts farther from the steady state than $F$ in both metrics (Appendix~\ref{app:controls}).

\section{Local measurement of asymptotic relaxation rates}

\begin{figure*}[!tp]
\includegraphics[width=.98\textwidth]{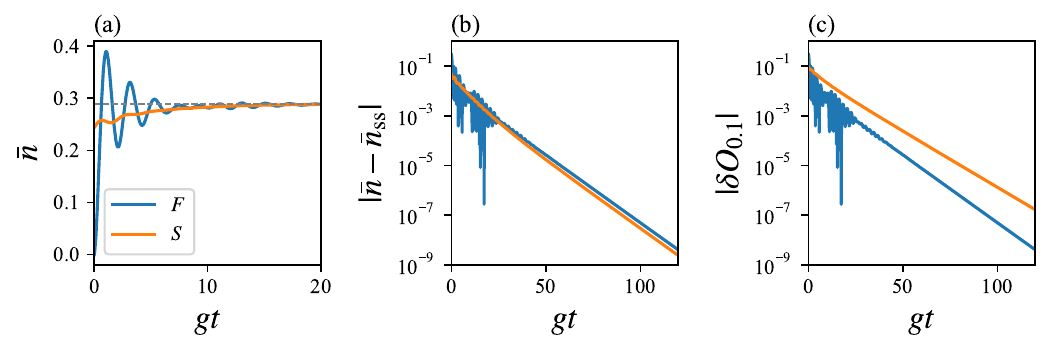}
\caption{Local measurement of asymptotic relaxation rates at $L=18$, $\kappa/g=0.1$, and $\beta g=1$. (a) Mean excitation density and its stationary value $\bar n_{\rm ss}=0.2894$ (dashed line). (b) Absolute error in the excitation density. At $\epsilon=10^{-3}$, the settling times are $23.1/g$ for $F$ and $21.0/g$ for $S$. (c) Absolute error $|\delta O_{0.1}|$ in $O_a$ at $a=0.1$, revealing the asymptotic rate ordering $\Gamma_F>\Gamma_S$. The initial errors are $0.2894$ for $F$ and $0.0889$ for $S$. Dips occur when an oscillating signal passes near its stationary value.}\label{fig:prep}
\end{figure*}

The excitation density is even under $\Theta$ and couples only to even decay modes [Fig.~\ref{fig:prep}(b)]. A local readout of the Gibbs rate $\Gamma_S$ needs overlap with the slow odd mode~\cite{BaguiObservables2026}.

The energy $H$ is odd under $\Theta$ and provides a complementary observable accessible through single-site measurements. With $P_{\mathcal B}$ the projector onto the blockade space, $P_{\mathcal B}\sigma_j^xP_{\mathcal B}=P_{\mathcal B}P_{j-1}\sigma_j^xP_{j+1}P_{\mathcal B}$, since a flip beside an excited neighbor takes a configuration outside this space. Thus $\langle H\rangle/(gL)=L^{-1}\sum_j\langle\sigma_j^x\rangle$ for states supported on the blockade subspace. To access the dominant decay modes of both preparations through the same observable, we combine the mean excitation density with the transverse spin,
\begin{equation}
	O_a=\frac{N_r}{L}+a\frac{H}{gL}.
	\label{eq:witness}
\end{equation}
Since the steady state is even, $\langle O_a\rangle_{\rm ss}=\bar n_{\rm ss}$. For the all-ground preparation, the transverse spin vanishes throughout the evolution, so the first term in Eq.~\eqref{eq:witness} yields the asymptotic relaxation rate $\Gamma_F$. For the Gibbs reference, the second term overlaps with the slow odd mode and dominates the late-time decay for $a\ne0$, yielding $\Gamma_S$. Figure~\ref{fig:prep}(c) shows that the local readout reproduces the rate ordering $\Gamma_F>\Gamma_S$.

The two terms in $O_a$ can be measured on separately prepared arrays through direct site-resolved imaging and a calibrated basis rotation before detection, respectively. Site-resolved imaging provides access to Rydberg populations and spatial correlations~\cite{Schauss2012,Ebadi2021,Scholl2021}, while single-site ground--Rydberg coherence has also been measured experimentally~\cite{XiangLocal2026}. Measuring the squared Hilbert--Schmidt distance requires the state purities and their overlap, accessible in principle through two-copy or randomized measurement protocols~\cite{Islam2015,Brydges2019}.

The all-ground array overshoots the steady-state density and approaches it through damped oscillations [Fig.~\ref{fig:prep}(a)]. Writing the instantaneous readout deviation as $\delta O_a(t)=\langle O_a\rangle_t-\langle O_a\rangle_{\rm ss}$, we define the settling time as the last entrance into an absolute tolerance band,
\begin{equation}
	\tau_{O_a}(\epsilon)=\inf\{t:|\delta O_a(u)|\le\epsilon
	\ \text{for all }u\ge t\}.
\end{equation}
Using the last crossing accounts for oscillations~\cite{Chatterjee2024,XiaMulti2026}. The asymptotic rate governs the settling time at small tolerances. At fixed readout variance, the number of independent shots required for precision $\epsilon$ scales as $\epsilon^{-2}$. The variance of a spatial average depends on intersite correlations. For independent sites and unit single-site variance, a standard error of $10^{-4}$ at $L=18$ requires about $6\times10^6$ shots.

Although the Gibbs density settles earlier in Fig.~\ref{fig:prep}(b), the combined readout settles earlier for the all-ground preparation at all tolerances in Table~\ref{tab:settling}. This ordering persists over a range of transverse-spin weights at percent-level precision. Numerical details are given in Appendix~\ref{app:num}.

\begin{table}[t]
	\caption{Settling times measured from pump onset for the all-ground and Gibbs preparations at $L=18$, $a=0.1$, and $\kappa/g=0.1$. The tolerance is an absolute error in $\langle O_a\rangle$.}\label{tab:settling}
	\begin{ruledtabular}
		\begin{tabular}{ccc}
			$\epsilon$ & $g\tau_{O_a,F}$ & $g\tau_{O_a,S}$\\
			\hline
			$10^{-2}$ & 6.9 & 16.0\\
			$10^{-3}$ & 23.1 & 36.6\\
			$10^{-4}$ & 39.5 & 58.4\\
		\end{tabular}
	\end{ruledtabular}
\end{table}

On the eight-site ring, detuning mixes the even and odd sectors. Both preparations decay with the same exponent at the nonzero detunings tested. The all-ground readout has a smaller slow-mode amplitude and settles earlier at the tolerances in Table~\ref{tab:detuning}.

The all-ground preparation starts from the ground-state arrays routinely used in atom assembly~\cite{Endres2016,Barredo2016}. If only excitation counting is available, a different initial mixture of the all-ground and alternating configurations can suppress a density-visible mode (Appendix~\ref{app:mixture}).

\section{Preparing and detecting the slow coherence}
\label{sec:pulse}

Before turning on the common drive and pump on an even-length ring, we choose the even or odd sublattice with equal probability and apply a parallel $\pi/2$ pulse to its atoms. The other atoms remain in $|g\rangle$, so each realization respects the blockade. With the pulse phase fixed relative to the coherent drive, the addressed atoms acquire the state $|u_s\rangle=(|g\rangle+s|r\rangle)/\sqrt2$ with $s\in\{+1,-1\}$, where shifting the phase by $\pi$ reverses $s$. Writing $|A_s\rangle=|u_s\,g\,u_s\,g\cdots\rangle$ and denoting one-site translation by $T$, we obtain the state at pump onset,
\begin{equation}
 \rho_{P_s}=\tfrac12\left(|A_s\rangle\langle A_s|+
              T|A_s\rangle\langle A_s|T^\dagger\right).
 \label{eq:pulsepreparation}
\end{equation}
Site-dependent light shifts have been combined with coherent Rydberg driving in neutral-atom arrays~\cite{OmranLocal2019}. The preparation requires sublattice selectivity and a pulse short enough to limit interaction phases between addressed atoms.

\begin{figure*}[tp]
\includegraphics[width=.98\textwidth]{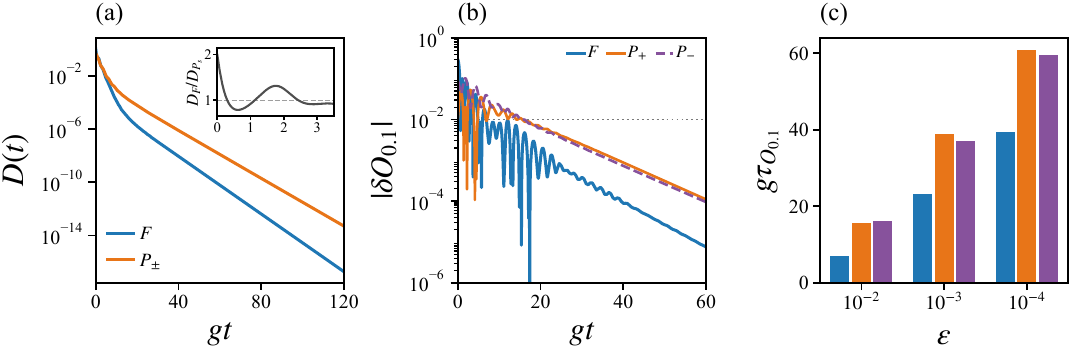}
\caption{Strong Mpemba relaxation with a pulsed reference at $L=18$ and $\kappa/g=0.1$. (a) Squared Hilbert--Schmidt distances for $F$ and $P_\pm$. Both phases give the same orange curve. The inset resolves the early crossings of $D_F/D_{P_s}$ through unity (dashed line). (b) Absolute error $|\delta O_{0.1}|$ in $O_a$ for $a=0.1$. (c) Settling times $g\tau_{O_{0.1}}$ at three tolerances. At the tolerance $0.01$ marked in (b), $g\tau_{O_{0.1}}$ is $6.9$ for $F$ (blue), $15.6$ for $P_+$ (orange), and $16.0$ for $P_-$ (purple). All times are measured from the onset of pumping.}\label{fig:pulse}
\end{figure*}

Mixing the two sublattices lowers the purity to near $1/2$. At $L=18$ the mixture is closer to the steady state than the all-ground state in both squared Hilbert--Schmidt and trace distance. The trace-distance margin is $3.85\times10^{-5}$ (Appendix~\ref{app:pulse}). The pulse populates the same slow odd mode as the Gibbs reference. The all-ground array remains even, overtakes both pulse phases after several early crossings, and decays at the faster even-sector rate [Fig.~\ref{fig:pulse}(a)].

Since $\Theta(\rho_{P_+})=\rho_{P_-}$, both arrays have the same excitation density and distance to stationarity at every time. Their transverse spins have opposite signs. Subtracting the two local readouts therefore isolates the coherence that carries the slow decay,
\begin{equation}
	\frac{\langle O_a\rangle_{P_+}-\langle O_a\rangle_{P_-}}{2a}
	=\frac{1}{L}\sum_j\langle\sigma_j^x\rangle_{P_+},\qquad a\ne0.
	\label{eq:phase_difference}
\end{equation}
The decay rate of this difference agrees with the odd-sector eigenvalue. In the combined readout, the all-ground array reaches and stays within the chosen tolerance earlier than either pulsed array [Figs.~\ref{fig:pulse}(b) and \ref{fig:pulse}(c)]. Using the excitation density alone can reverse this settling-time ordering (Appendix~\ref{app:pulse}). In practice, the pulsed preparations require additional time for sublattice addressing and a $\pi/2$ pulse of duration $\pi/(2\Omega_p)$ for a square pulse with Rabi frequency $\Omega_p$.

\begin{figure*}[!t]
\includegraphics[width=.96\textwidth]{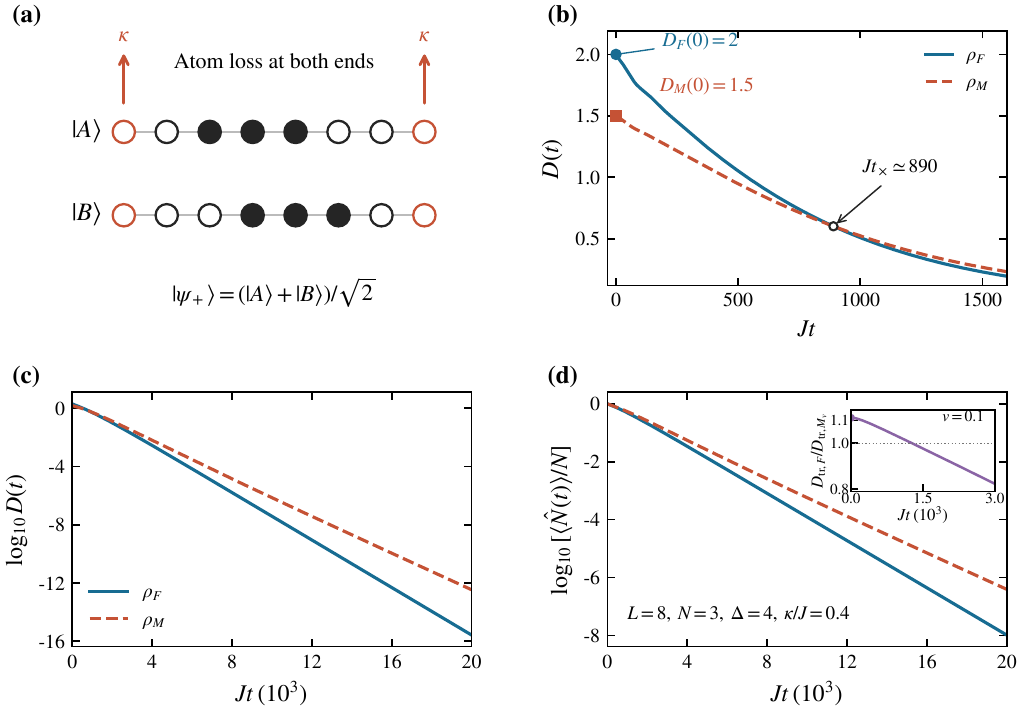}
\caption{Endpoint loss in the XXZ chain, $L=8$, initial $N=3$, $\Delta=4$, and $\kappa/J=0.4$. (a) Reflected configurations $|A\rangle=|00111000\rangle$ and $|B\rangle=|00011100\rangle$ and endpoint loss in the hard-core representation. (b,c) Early and late squared Hilbert--Schmidt distance dynamics for the even coherent preparation $\rho_F$ and incoherent mixture $\rho_M$. Their initial values are $2$ and $3/2$, while both initial trace distances equal one. (d) Decay of $\langle\hat N(t)\rangle/N$, where $\hat N=\sum_{j=1}^L n_j$ and $N=3$ is the initial excitation number of $F$ and $M$. The inset compares trace distances for $\rho_F$ and $\rho_{M_v}=(1-v)\rho_M+v\rho_{\rm vac}$, with vacuum weight $v=0.1$. The admixture gives $\rho_{M_v}$ a fluctuating initial number. The decay rates, including all populated number sectors, are $\Gamma_F/J=9.395\times10^{-4}$ and $\Gamma_M/J=7.188\times10^{-4}$.}\label{fig:xxz}
\end{figure*}

\section{Spin-flip loss in a magnetization-conserving XXZ chain}

The Hamiltonian of the open XXZ chain is
\begin{equation}
 H_{\rm XXZ}=J\sum_{j=1}^{L-1}
 \left(S_j^xS_{j+1}^x+S_j^yS_{j+1}^y+
 \Delta S_j^zS_{j+1}^z\right).
\end{equation}
The Hamiltonian conserves $S_{\rm tot}^z$. Independent flips at the endpoints, $J_1=\sqrt\kappa S_1^-$ and $J_L=\sqrt\kappa S_L^-$, give
\begin{equation}
 \mathcal L^\dagger(S_{\rm tot}^z)=-\kappa(n_1+n_L),
\end{equation}
where $n_j=S_j^z+1/2$. Starting in a fixed magnetization sector, each jump removes one excitation. Hopping can bring the remaining excitations to either end, and the chain relaxes to the fully down-polarized vacuum $\rho_{\rm vac}=|0\cdots0\rangle\langle0\cdots0|$. We write $\hat N=\sum_{j=1}^L n_j$ for the excitation-number operator and $N$ for the initial excitation number of a fixed-number state. This sequence of coupled number sectors also appears in exact treatments of quantum dynamics under loss~\cite{Buca2020}. Uniform single-site loss gives identical trace-distance curves for all states with the same fixed $N$ (Appendix~\ref{app:xxz_uniform}). Endpoint loss resolves reflection-related cluster modes.

Figure~\ref{fig:xxz}(a) shows two configurations $|A\rangle$ and $|B\rangle$ of the same compact cluster, related by reflection about the center of the chain. They define a coherent preparation $\rho_F=|\psi_+\rangle\langle\psi_+|$ with $|\psi_+\rangle=(|A\rangle+|B\rangle)/\sqrt2$ and a mixture $\rho_M=(|A\rangle\langle A|+|B\rangle\langle B|)/2$. The excitation number, site populations, and mean energy are the same in both. Compact clusters form narrow bands in the easy-axis XXZ spectrum~\cite{Nachtergaele2001,Nachtergaele2007,Fischbacher2014}. The mixture populates both reflection parities of the cluster. The coherent state occupies only the even superposition and therefore avoids a slowly leaking cluster mode in its initial number sector. In this example, the compact cluster lies away from the loss sites.

An endpoint flip can populate slower modes in sectors with fewer excitations. Including all sectors reached under loss, the coherent state still relaxes faster in the finite chain of Fig.~\ref{fig:xxz}. It starts farther from the vacuum in squared Hilbert--Schmidt distance and overtakes the mixture [panels (b) and (c)]. Both fixed-number preparations initially have unit trace distance from the vacuum, because their supports are orthogonal to it. A small vacuum admixture, $\rho_{M_v}=(1-v)\rho_M+v\rho_{\rm vac}$ with $v=0.1$, gives strict initial ordering in trace distance while preserving its decay exponent. The resulting trace-distance crossing is shown in the inset of panel (d).

In the hard-core representation, $S_j^-=b_j$ and an excitation denotes an occupied site. Up to a uniform chemical-potential term and a constant, the atomic Hamiltonian is
\begin{align}
 H_{\rm at}={}&\frac J2\sum_{j=1}^{L-1}(b_j^\dagger b_{j+1}+\mathrm{H.c.})
 +J\Delta\sum_{j=1}^{L-1}n_jn_{j+1}\nonumber\\
 &+\frac{J\Delta}{2}(n_1+n_L).
\end{align}
The edge potential follows from the open boundary conditions. Under the mapping $S_j^-=b_j$, the jump operators become $J_1=\sqrt{\kappa}b_1$ and $J_L=\sqrt{\kappa}b_L$, describing single-particle loss at both ends of the atomic chain. Local one-body loss~\cite{Barontini2013,Labouvie2016} and interaction-dependent two-body loss~\cite{Tomita2017,Tomita2019} have been realized in cold-atom experiments.

The resonant pump preserves the $\Theta$ sector of the Rydberg density matrix throughout the evolution. In the XXZ chain, reflection parity restricts the initial cluster modes, but each endpoint loss opens a lower-number sector. The rate advantage survives as long as the modes reached along this sequence all decay faster than the slow reference mode.

\FloatBarrier
\section{Conclusion and outlook}

At $\kappa/g=0.1$ the all-ground array outpaces the Gibbs reference at $\beta g=1$ and both pulsed references in preparing the correlated Rydberg steady state. It stays in the even sector, so the slow odd mode populated by the references never enters its evolution. Independent driven atoms already exhibit this mode selection. The blockade gives the target its spatial correlations.

Opposite pulse phases yield the same population dynamics and opposite transverse-spin signals. The difference of the two readouts isolates the slow coherence. For the Gibbs comparison on the detuned eight-site ring, both preparations share the same late-time rate. The smaller slow-mode signal of the all-ground array still shortens its settling time at the tolerances studied.

An experiment needs a pump that resolves the blockade shift and a phase-calibrated spin rotation before site-resolved imaging~\cite{Schafer2020,Weimer2010}. Measurements on driven clusters and controlled dissipation characterize the relevant shifts and decay channels~\cite{Ding2024,Begoc2025,ChenRydberg2025}. Measuring the pump rate beside an excited neighbor would quantify leakage from the blockaded space.

\begin{acknowledgments}
This work was supported by the National Natural Science Foundation of China under Grant Nos.~12304290, 12505017 and 12674355, and by the Beijing National Laboratory for Condensed Matter Physics under Grant No.~\mbox{2025BNLCMPKF017}. L.P. also acknowledges support from the Fundamental Research Funds for the Central Universities (010-63263115).
\end{acknowledgments}
\appendix
\section{Numerical construction and verification}\label{app:num}

The $L=18$ ring has $d_L=5778$ allowed binary configurations. Writing $c_j=J_j/\sqrt\kappa$ and $F_j=c_j^\dagger c_j$, column vectorization gives
\begin{align}
 \mathbb L={}&-i(I\otimes H-H^T\otimes I)\\
 &+\kappa\sum_j\left[c_j^*\otimes c_j
 -\tfrac12(I\otimes F_j+F_j^T\otimes I)\right].\nonumber
\end{align}
We sum configuration dyads over translations and reflections. Combining each orbit with its transpose gives orthonormal Hermitian bases of definite $\Theta$ parity. The even and odd blocks at $L=18$ have dimensions $468118$ and $464905$. Together they span the spatial symmetry sector of the preparations.

We diagonalize the complete accessible blocks through $L=12$ and use sparse Arnoldi iterations to find their leading modes at larger $L$~\cite{Lehoucq1998}. Left eigenvectors give the projections of each initial density matrix. At $L=18$, the magnitudes of the leading even projection of the all-ground state and the leading odd projection of the Gibbs reference are $0.00893$ and $0.0310$ in the numerical normalization. The all-ground odd projection vanishes by symmetry. We cross-check the direct and adjoint mode calculations at $L=14$ and $18$ by applying Arnoldi iterations to the finite-time propagator on the trace-zero subspace. At $L=18$ its action is evaluated with a Chebyshev expansion~\cite{TalEzer1984}. The two methods give decay rates within $4\times10^{-12}g$ of each other, with right-mode residuals below $10^{-10}g$ in the original generator. At $L=16$ we use the direct and adjoint generator calculations.

We obtain the initial trace distances by summing the trace norms of the density-matrix deviations over their physical Hilbert-space momentum blocks and dividing by two. The same decomposition gives a positive minimum eigenvalue for the trace-one $L=18$ stationary state. Translation-orbit matrix elements give its excitation distribution and diagonal correlations. The residual in the stationary population balance derived in Appendix~\ref{app:steady} is $2.1\times10^{-17}g$.

For the time evolution we propagate $\delta\rho_i=\rho_i-\rho_{\rm ss}$ directly. The squared Hilbert--Schmidt distance is the sum of the squared coefficients in the orthonormal Hermitian basis, including both parities for the Gibbs reference. Chebyshev exponential actions evolve each deviation to $gt=120$. Observables are sampled every $g\Delta t=0.05$ up to $gt=24$, and all distances and observables every $g\Delta t=0.5$ throughout. The two observable grids agree to $1.2\times10^{-15}$ at coincident times. An independent exponential action~\cite{AlMohy2011} at $gt=2$ gives relative state-vector differences below $7.2\times10^{-15}$. Replacing one time step by two half steps gives a relative change below $1.3\times10^{-15}$. The largest trace drift is $2.7\times10^{-16}$. Sparse spectral calculations give $\Gamma_F/g=0.12489$ and $\Gamma_S/g=0.10310$, while fits to $\log D_i(t)$ over $100\le gt\le120$ yield $\Gamma_F/g=0.12484$ and $\Gamma_S/g=0.10311$. 

Each settling time in the main text is estimated by interpolating the last sampled crossing of the indicated tolerance. To control the unobserved tail, take $s\ge T$ and write the observable error as $\operatorname{Tr}[(O_a-cI)\delta\rho_i(s)]$ for any real $c$. Trace-norm contraction bounds $\|\delta\rho_i(s)\|_1$ by $\|\delta\rho_i(T)\|_1$. Hence
\begin{equation}
 |\langle O_a\rangle_s-\langle O_a\rangle_{\rm ss}|\le
 \|O_a-cI\|_\infty\sqrt{d_L D_i(T)}.
\end{equation}
At $c=1/4$, Eq.~\eqref{eq:witness} gives $\|O_a-cI\|_\infty\le1/4+|a|$ since $0\le N_r/L\le1/2$ and $\|H\|_\infty\le gL$. For $O_{0.1}$ this gives $\|O_{0.1}-cI\|_\infty\le0.35$. For $gT=120$, the $O_{0.1}$ error therefore stays below $1.18\times10^{-7}$ for $F$ and $4.70\times10^{-6}$ for $S$ at all later times. For $t<T$, the resolution of the last crossing is set by the sampling interval.

In Fig.~\ref{fig:dyn}, the initial trace distances are $0.99991$ for $F$ and $0.88676$ for $S$. Varying the spin weight in Eq.~\eqref{eq:witness} gives Gibbs settling times of $12.1/g$ and $21.0/g$ for $a=0.05$ and $0.20$ at $\epsilon=0.01$. Both endpoint signals are still below their stationary value by more than $0.01$ at $gt=10$. The signal is affine in $a$, so this inequality holds throughout the interval of weights. The all-ground settling time remains $6.9/g$ because its transverse spin vanishes.

\section{Stationary coherence and observable selection}\label{app:steady}

Consider a density matrix with diagonal probabilities $p_a$ in the configuration basis. Each $c_j$ maps a configuration to at most one other configuration, so the dissipator preserves diagonality. The off-diagonal stationary equations would require $H_{ab}(p_b-p_a)=0$ on every allowed flip edge. Each configuration connects to the vacuum by removing its excitations. The probabilities would therefore be uniform over the entire connected graph. This gives the vacuum an outflow $\kappa Lp_{\rm vac}$ with zero incoming pump rate. For $g,\kappa>0$, stationarity consequently requires at least one off-diagonal matrix element.

On Hermitian operators, $\operatorname{Tr}[\Theta(A)\Theta(B)]=\operatorname{Tr}(AB)$. Opposite parities thus have zero trace pairing. Every configuration-diagonal observable, including $N_r$, couples only to even modes. The energy $H$ is odd and probes the odd component of the Gibbs reference. For $\|A\|_\infty\le1$, the expectation error $|\operatorname{Tr}[A(\rho-\rho_{\rm ss})]|$ gives a lower bound on $2D_{\rm tr}$.

Defining $Y_j^{(c)}=i(c_j-c_j^\dagger)$, the master equation gives the excitation balance
\begin{equation}
 \frac{d\langle N_r\rangle}{dt}=-g\sum_j\langle Y_j^{(c)}\rangle+\kappa\sum_j\langle F_j\rangle .
\end{equation}
Here $F_j$ projects onto configurations with sites $j-1,j,j+1$ all unexcited. In the translation-invariant stationary state, $\langle Y_j^{(c)}\rangle=(\kappa/g)f_{\rm ss}$, where $f_{\rm ss}=\langle F_j\rangle=1-3\bar n_{\rm ss}+\langle n_jn_{j+2}\rangle_{\rm ss}$. Within the blockaded Hilbert space $\langle\sigma_j^+\rangle=\langle c_j\rangle$, and $\Theta$ evenness fixes $\langle\sigma_j^x\rangle_{\rm ss}=0$. The reduced density matrix in the ordered basis $(|g\rangle,|r\rangle)$ is
\begin{equation}
 \rho_{\rm ss}^{(1)}=\begin{pmatrix}1-\bar n_{\rm ss}&-i\kappa f_{\rm ss}/(2g)\\i\kappa f_{\rm ss}/(2g)&\bar n_{\rm ss}\end{pmatrix}.
\end{equation}
At $L=18$ we find $\bar n_{\rm ss}=0.28936$, $f_{\rm ss}=0.25678$ and $\langle n_jn_{j+2}\rangle_{\rm ss}=0.12486$, giving $\langle g|\rho_{\rm ss}^{(1)}|r\rangle=-0.01284i$. The full state has purity $2.533\times10^{-4}$, above the maximally mixed value $1/5778$. Neighboring excitations have a negative connected correlation $-\bar n_{\rm ss}^2$, since $n_jn_{j+1}=0$ throughout the constrained space.

\section{Uniform-loss comparison for the XXZ chain}\label{app:xxz_uniform}

For a hard-core chain with $[H_{\rm at},\hat N]=0$ and equal lowering rates on all sites, $\sum_jJ_j^\dagger J_j=\kappa \hat N$. Tracing the master equation over each $n$-particle subspace gives its probability $P_n$,
\begin{equation}
 \dot P_n=\kappa[(n+1)P_{n+1}-nP_n].
\end{equation}
For initial number $N_0$, the generating function $\sum_nP_n(t)z^n=[1-e^{-\kappa t}+ze^{-\kappa t}]^{N_0}$ yields $\langle \hat N\rangle=N_0e^{-\kappa t}$ and $P_0=(1-e^{-\kappa t})^{N_0}$. An initial state in a single number sector remains block diagonal in $\hat N$, so its trace distance from the vacuum equals $1-P_0$. All preparations with the same fixed $N_0>0$ then have identical trace-distance curves and long-time exponent $\kappa$. Endpoint loss weights the spatial distribution at the boundary and resolves the cluster modes of Fig.~\ref{fig:xxz}~\cite{Buca2020,Barontini2013,Labouvie2016}.

\begin{figure}[!t]
\includegraphics[width=\columnwidth]{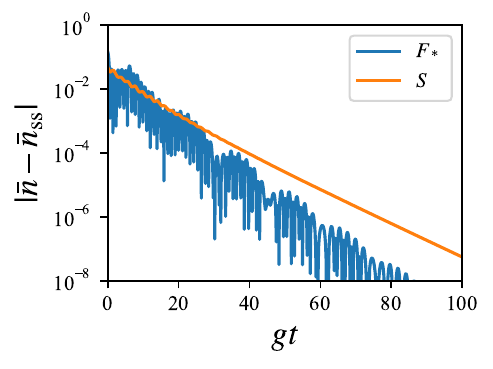}
\caption{Relaxation of the excitation density in a ten-site array with $\kappa/g=0.1$. At $p=p_*\simeq0.292974$, the classical mixture $F_*$ has zero overlap with the leading density-visible mode. The orange curve ($S$) is the $\beta g=1$ Gibbs reference.}\label{fig:mixsmall}
\end{figure}

\section{Population relaxation in a ten-site array}\label{app:mixture}

The mean density couples to even modes of the generator. A mixture of all-ground and alternating configurations allows the leading density-visible coefficient to be varied continuously. Define
\begin{gather}
 \rho_{F_*}=(1-p)\rho_F+p\rho_{\rm alt},\\
 \rho_{\rm alt}=\tfrac12\sum_{a=1}^2|Z_2^{(a)}\rangle\langle Z_2^{(a)}|,
\end{gather}
where $|Z_2^{(1)}\rangle=|rgrg\cdots\rangle$ and $|Z_2^{(2)}\rangle=|grgr\cdots\rangle$. Both density matrices are diagonal in the configuration basis. Averaging the two alternating patterns restores translation invariance. The resulting mixture is $\Theta$ even for every $p$.

At $L=10$, the two components have opposite projections onto the leading even left mode. Its coefficient is linear in $p$ and vanishes at $p_*\simeq0.292974$. At this weight the mode with rate $0.119396g$ drops out, leaving a conjugate pair with decay rate $0.171158g$ and angular frequency $0.008338g$. Figure~\ref{fig:mixsmall} shows the density error relative to $\bar n_{\rm ss}=0.29226$. The mixture starts farther from this density than the Gibbs reference, with absolute errors $0.14577$ and $0.04855$. Their late-time envelopes give a rate ratio of $1.4335$.

At tolerance $10^{-5}$, the density settles at $43.75/g$ for the mixture and $57.50/g$ for the reference, a time reduction of $23.9\%$. A similar reduction occurs at $10^{-6}$. At coarser tolerances the early transient and the larger initial error can determine the ordering. Changing $p$ away from $p_*$ restores a component with rate $0.119396g$, although the preparations at $p_*\pm0.01$ retain a finite-tolerance advantage in the numerical tests.

\section{Independent atoms and weak perturbations}\label{app:controls}

\emph{Independent atoms.} Removing the blockade projectors from both $H$ and $J_j$ gives $L$ identical atoms with single-site Hamiltonian $g\sigma^x$ and jump $\sqrt\kappa\,\sigma^+$. Write $x=\langle\sigma^x\rangle$ and choose $z$ positive in $|r\rangle$. The $x$ deviation decays at $\kappa/2$. The other two Bloch deviations have eigenvalues $-3\kappa/4\pm\sqrt{\kappa^2/16-4g^2}$. In the underdamped regime $\kappa<8g$, both have decay rate $3\kappa/4$. The ground state has $x=0$ and the single-atom Gibbs state has $x=-\tanh(\beta g)$. Each many-body preparation evolves as a tensor product, and its expansion around the stationary product state contains terms linear in these single-site deviations. Products of two or more deviations decay faster. This gives $\eta=3/2$ at every finite $L$ for $\beta g>0$.

The initial squared distances also have the requisite ordering. The stationary Bloch vector has $x_{\rm ss}=0$ and $z_{\rm ss}=\kappa^2/(\kappa^2+8g^2)$. The single-site overlaps with the target are $b=(1-z_{\rm ss})/2$ for the ground state and $1/2$ for the Gibbs state. Denoting the Gibbs purity by $p_\beta=[1+\tanh^2(\beta g)]/2$, the difference of the two squared distances is $1-p_\beta^L+2(2^{-L}-b^L)>0$.

\begin{figure}[!t]
\includegraphics[width=\columnwidth]{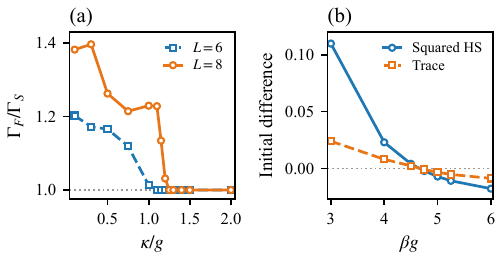}
\caption{Finite-ring parameter checks. (a) Populated rate ratio for $F$ and $S$ at $\beta g=1$. (b) Initial squared Hilbert--Schmidt and trace-distance differences $F-S$ at $L=8$ and $\kappa/g=0.1$. Symbols are calculated points and lines guide the eye.}\label{fig:scope}
\end{figure}

\emph{Perturbed blockaded ring.} To probe experimental tolerances we keep the full $L=8$ blockade basis ($d_L=47$) and change the Hamiltonian to $H_\delta=H-\delta N_r$, where $H$ denotes the resonant PXP drive. The initial Gibbs state and the readout $O_{0.1}$ both use $H$ throughout the scan. We recompute the steady state at each parameter point. The translation- and reflection-invariant operator space has dimension 187 and admits complete diagonalization. Uniform detuning mixes the two resonant $\Theta$ sectors. For every tested nonzero detuning, both preparations populate the same isolated slow mode, yielding $\eta=1$. At $\delta/g=0.05$, for example, its decay rate is $0.09445g$ and its amplitudes in the measured signal are $0.0020$ and $0.0279$ for $F$ and $S$. We extract the settling times from the full trajectories.

For $\delta/g=0,\pm0.01,\pm0.03,\pm0.05,\pm0.10$ with the upward pump alone, the reduction $1-\tau_{O_{0.1},F}/\tau_{O_{0.1},S}$ ranges from $13.6\%$ to $24.6\%$ at $\epsilon=0.01$ and from $10.4\%$ to $17.7\%$ at $\epsilon=0.001$. An additional downward jump $\sqrt{\kappa_\downarrow}\,P_{j-1}\sigma_j^-P_{j+1}$ preserves $\Theta$ at resonance. With $\kappa_\downarrow/g=0.01$ and $\kappa/g=0.1$, its two populated rates are $0.14570g$ and $0.10281g$, giving $\eta=1.42$. At $\delta/g=\pm0.05$ with both jumps the leading rates again coincide and $F$ retains the finite-tolerance advantage. Table~\ref{tab:detuning} lists all 12 cases.

\begin{table}[ht]
\caption{Settling times for the perturbed eight-site ring. All rows use $\kappa/g=0.1$ and the same initial pair. The two columns for each tolerance give $g\tau_{O_{0.1},F}$ and $g\tau_{O_{0.1},S}$ for the readout $O_{0.1}$.}\label{tab:detuning}
\begin{ruledtabular}
\begin{tabular}{rrrrrr}
$\delta/g$ & $\kappa_\downarrow/g$ & \multicolumn{2}{c}{$\epsilon=0.01$} & \multicolumn{2}{c}{$\epsilon=0.001$}\\
 & & $F$ & $S$ & $F$ & $S$\\
\hline
$-0.10$ & 0 & 14.2 & 18.8 & 33.5 & 40.7 \\
$-0.05$ & 0 & 14.2 & 17.9 & 33.4 & 39.9 \\
$-0.03$ & 0 & 14.2 & 17.7 & 33.4 & 39.5 \\
$-0.01$ & 0 & 14.2 & 17.6 & 33.3 & 39.2 \\
0 & 0 & 14.2 & 17.5 & 33.3 & 39.0 \\
$+0.01$ & 0 & 14.2 & 17.5 & 33.3 & 38.8 \\
$+0.03$ & 0 & 14.1 & 17.4 & 33.3 & 38.5 \\
$+0.05$ & 0 & 14.1 & 17.3 & 33.3 & 38.1 \\
$+0.10$ & 0 & 14.1 & 16.3 & 33.2 & 37.1 \\
$-0.05$ & 0.01 & 14.0 & 16.8 & 31.1 & 36.9 \\
0 & 0.01 & 14.0 & 16.1 & 29.1 & 36.2 \\
$+0.05$ & 0.01 & 14.0 & 15.7 & 25.8 & 35.4 \\
\end{tabular}
\end{ruledtabular}
\end{table}
The final crossing brackets have width $0.025/g$. We checked the later sampled intervals with a bound on the interpolation remainder, subdividing intervals near the threshold. Trace-norm contraction bounds the signal after $gT=120$ by $6.7\times10^{-6}$ across the scan.

\emph{Pump and temperature checks.} We diagonalize the translation- and reflection-invariant blocks at $L=6$ and $8$ and use left eigenvectors to check which modes are populated. At $\beta g=1$, the leading nonzero even and odd rates cross at $\kappa/g\simeq1.03$ for $L=6$ and $1.22$ for $L=8$. At the sampled pump strengths above each crossing, $F$, $S$, and $P_\pm$ populate the leading even mode [Fig.~\ref{fig:scope}(a)]. For $L=8$ and $\kappa/g=1.3$, its rate is $0.98743g$, compared with $1.13774g$ in the odd block. For $L=8$, the even rate times $\kappa/g^2$ is $0.8960$ and $0.8907$ at $\kappa/g=64$ and $128$, consistent with a Zeno scale $g^2/\kappa$. At $g=0$, configurations that admit no additional excitation are dark to the pump. A single coherent flip produces a coherence with a pump-active configuration. Two flips through these rapidly damped coherences generate the second-order scale.

At $\kappa/g=0.1$, the zeros of the initial squared Hilbert--Schmidt and trace-distance differences occur at $\beta g\simeq3.61$ and $3.62$ for $L=6$, and at $4.66$ and $4.67$ for $L=8$. For $L=8$ at $\beta g=5$, for instance, $D_F(0)-D_S(0)=-0.00706$ and $D_{{\rm tr},F}(0)-D_{{\rm tr},S}(0)=-0.00335$ [Fig.~\ref{fig:scope}(b)].

\section{Sublattice preparations and phase reversal}
\label{app:pulse}

For a pulse area $\theta$ and phase $\phi$, the selected-site spinor is $c|g\rangle+e^{i\phi}\sin(\theta/2)|r\rangle$, with $c=\cos(\theta/2)$. The translated branches overlap by $c^L$, so their equal mixture has eigenvalues $(1\pm c^L)/2$ and purity $(1+c^{2L})/2$. The transformation $\Theta$ sends $\phi$ to $\pi-\phi$. For $m=L/2$, the squared norm of the odd part of the initial density operator is
\begin{equation}
 \big\|\rho^{(-)}_{\phi}\big\|_2^2=
 \tfrac14\left[1-(1-\sin^2\!\theta\cos^2\!\phi)^m\right].
\end{equation}
In $\tfrac12\operatorname{Tr}[\rho_{\phi}^2-\rho_{\phi}\,\Theta(\rho_{\phi})]$, cross terms between sublattices cancel. A real $\pi/2$ pulse has odd squared norm $1/4$, while a pulse producing a purely imaginary relative amplitude is $\Theta$ even. At $L=18$, each real pulse preparation starts with $\bar n=1/4$ and $\langle H\rangle/(gL)=s/2$. Its overlap with the normalized left eigenoperator of the leading odd mode has magnitude $0.0400$. In a mixture $p\rho_{P_+}+(1-p)\rho_{P_-}$ the odd amplitude acquires a factor $2p-1$. Only equal mixing of pulse signs cancels the odd mode. Convexity of both initial distance measures ensures that the whole sign-mixture family lies initially closer than $F$ at the parameters used here.

The sum and difference of the translated branches have momenta $0$ and $\pi$. All other momentum blocks of $\rho_{P_s}-\rho_{\rm ss}$ are negative semidefinite. The trace distance is the sum of the positive eigenvalues in the two occupied blocks (dimensions 329 and 319). For Fig.~\ref{fig:pulse} we evolve both $\Theta$ sectors on a distance grid $g\Delta t=0.5$, refined to $0.025$ over $0\le gt\le6$. Readouts use spacing $0.05$ up to $gt=120$. An independent exponential action at $gt=2$ agrees within relative error $5.4\times10^{-15}$ in both sectors. Beyond $gt=120$ all pulse readout errors stay below $6.06\times10^{-6}$ by Appendix~\ref{app:num}. We interpolate crossing times within the sampled brackets.

Initial squared distances at $L=18$ are $1.00007$ for $F$ and $0.50013$ for either pulse phase. Translation invariance makes the steady-state overlap of either pure sublattice branch equal to the overlap of their mixture. The squared distance of a pure branch therefore exceeds the mixture value by the purity difference $(1-2^{-L})/2$. At $L=18$ this gives $1.000125$ for a single branch, slightly above $1.000074$ for $F$. Their initial trace distances are $0.99991017$ and $0.99987168$, respectively. At $\kappa/g=0.1$, the trace-distance margins for the sublattice mixture are $4.04\times10^{-2}$ at $L=6$, $9.81\times10^{-3}$ at $L=8$, and $3.85\times10^{-5}$ at $L=18$. The distances cross at $gt\simeq0.30,1.13,2.43$. At $gt=120$, $D_F/D_{P_s}=3.78\times10^{-4}$. Fits on $100\le gt\le120$ give $\Gamma_F/g=0.12484$ and $\Gamma_{P_s}/g=0.10312$, in agreement with the spectral ratio $\eta_{18}=1.21$. Over the same interval the phase difference of the readouts decays at $0.10310g$.

For $O_{0.1}$ the initial absolute errors are $0.28936$, $0.01064$, and $0.08936$ for $F$, $P_+$, and $P_-$, respectively. The $P_+$ readout starts close to the $0.01$ tolerance and undergoes an oscillatory transient. At $\epsilon=10^{-3}$ the three settling times are $23.07/g$, $38.74/g$, and $37.00/g$. For excitation density alone, both pulse preparations settle at $14.50/g$ at this tolerance, ahead of the all-ground density at $23.07/g$.

\onecolumngrid
\clearpage
\twocolumngrid
\bibliographystyle{apsrev4-2}
\makeatletter\immediate\write\@auxout{\string\citation{prbBibliographyControl}}\makeatother
\bibliography{references}
\end{document}